# Reversible grain boundary complexion transition explains cratonic lithosphere anomalies

Alexandra C. Austin[1,2*], Sanae Koizumi[3], Martin Folwarczny[1,2], David P. Dobson[4], Katharina Marquardt[1,2,5]

[1]Imperial College London, London, United Kingdom
[2]University of Oxford, Oxford, United Kingdom
[3]National Institute for Materials Science: Tsukuba, Ibaraki, Japan
[4]University College London, London, United Kingdom
[5]The Zero Institute, Oxford, OX2 0ES, UK

*Alexandra Austin

**Email:** alexandra.austin@materials.ox.ac.uk

**Author Contributions: A.C.A.** methodology, validation, formal analysis, investigation, data curation, writing, reviewing and editing, visualization. **S. K.** Resources, reviewing and editing. **M. F.** Software, reviewing and editing. **D. P. D.** conceptualization, validation, resources, reviewing and editing, supervision, funding acquisition. **K. M.** conceptualization, methodology, validation, resources, reviewing and editing, visualization, supervision, project administration, funding acquisition.

**Competing Interest Statement:** The authors declare no competing interests.

### Abstract

Seismic, electrical, and rheological properties of the Earth's lithosphere change abruptly at the lithosphere-asthenosphere boundary (LAB), but the underlying mechanism driving this in dry cratonic keels remains unresolved. Here we present experimental evidence that grain boundaries in forsterite ($Mg_2SiO_4$) undergo a reversible complexion transition between 1290 °C and 1390 °C. Using electron backscatter diffraction (EBSD) and stereological analysis, we show that the grain boundary plane distribution (GBPD) shifts from {0kl}-type (*c-type*) to {010} (*b-type*) grain boundaries across this narrow temperature interval, independent of grain size and impurities. This is the first direct experimental observation of a reversible complexion transition in a rock-forming mineral. When scaled to mantle geotherms, the transition coincides with depths of ~180-200 km where conductivity and viscosity discontinuities are detected beneath stable cratons. Our results demonstrate that grain boundary complexion transitions can influence geophysical observables, offering a new framework for interpreting lithosphere-asthenosphere coupling in the absence of melt or hydration.

### Summary paragraph

Sharp changes in seismic velocity, viscosity, and electrical conductivity beneath stable cratons remain unexplained in the absence of melt or dissolved hydrogen. Studies of olivine, the most abundant mineral in the upper mantle, can inform interpretations of geophysical observables from this region. Geophysical observations reveal a sharp decrease in viscosity and a concurrent increase in conductivity across the lithosphere-asthenosphere boundary, occurring at depths of 150-230 km, corresponding to temperatures near 1300 °C. Here we show that olivine undergoes a reversible grain boundary complexion transition between 1290 °C and 1390 °C. This transition alters the lowest-energy grain boundary plane and the physical properties of polycrystalline aggregates. Because grain boundaries control both mechanical strength and electrical transport, this transition offers a mechanistic explanation for the observed changes in viscosity and conductivity beneath old cratonic keels such as the Slave and Kaapvaal regions. This is the first experimental evidence of a complexion transition in olivine, offering insight into depth-dependent rock properties and resolving inconsistencies in lab measurements.

**Introduction**

The 1300 °C geotherm, often used to define the lithosphere-asthenosphere boundary (LAB), occurs at ~60-120 km beneath oceanic plates and up to ~250 km beneath continental cratons. Across this boundary, geophysical observables change markedly: seismic shear velocities decrease, seismic attenuation rises, electrical conductivity increases, and viscosity drops by several orders of magnitude (Rychert et al., 2020). These shifts are attributed to increasing temperatures, where olivine begins to weaken dramatically, producing a transition from elastic, rigid behaviour to ductile creep (Bystricky et al., 2024; Hansen et al., 2019). Elevated temperatures also allow trace amounts of melt or hydrated defects, which enhance conductivity, decrease strength and further lower seismic velocities (Chen et al., 1998; Karato et al., 1986; Liu et al., 2023; McCarthy & Takei, 2011). Above the LAB, heat is mainly transported by conduction, sustaining a cold, strong lithosphere; below, convection dominates, facilitating mantle flow. In upwelling regions, the 1300 °C isotherm approaches the dry peridotite solidus, allowing incipient partial melting that further amplifies the seismic and conductive anomalies (Hirschmann, 2000). Together, these processes explain why the LAB is expressed across multiple geophysical observables. Yet, beneath stable cratons, where melt and water are absent, electrical conductivity and viscosity change surprisingly abruptly and over a narrow depth range; both the narrow depth range and the abruptness remain unexplained.

In dry cratonic keels such as beneath the Kaapvaal, Slave, and Siberian cratons, seismic shear velocities remain high (>4.7 km/s) and high resistivity persists to ~200-250 km, reflecting buoyant, Fe-depleted, Mg-rich peridotite that is chemically dry (Baptiste & Tommasi, 2014; Bettac et al., 2023a; Evans et al., 2011). The lack of water and melt suppresses conductivity and maintains high seismic velocities (J. Chen et al., 1998; Jackson et al., 2002; Liu et al., 2023), while olivine grains with low defect densities sustain strong lattice-preferred orientation and high anisotropy (Ben Ismaı ̈l & Mainprice, 1998; Miyazaki et al., 2013). Dehydration and melt extraction during Archean formation further removed weakening agents, producing a rheologically strong keel (>$10^{22}$ Pa·s) (Bedle et al., 2021; King, 2005).Yet, distinct geophysical boundaries reveal where this strength begins to fade. The magnetotelluric boundary, estimated to be between 210-250 km deep below continents (at T=1200-1300 °C) (Jones et al., 2010), marks a drop in resistivity. Likewise, the 1300 °C isotherm defines a rheological threshold: at lower temperatures, dislocation creep is limited, and the lithosphere remains stiff (Mei et al., 2010), but above this temperature, deeper in the mantle, diffusion and grain-boundary sliding become more effective, enabling gradual weakening even in nominally dry peridotite (Bollinger et al., 2019; Raterron et al., 2019; Yabe et al., 2020). Together, these transitions delineate the cratonic LAB where rigid, depleted keels give way to warmer, weaker asthenosphere despite their extreme dryness.

Much work has been done to understand rheological changes at the LAB, but two points remain unexplained by experiments. Firstly, there is an abrupt decrease in viscosity around 210-250 km, corresponding to temperatures of ~1200-1300 °C below old, dry cratons (Xu et al., 2025). Notably, the magnitude of the decrease in viscosity is large ($10^{21}$-$10^{19}$ Pa·s) but occurs over a small change in depth (relatively constrained temperature and pressure conditions). This remains unexplained by classical flow laws for olivine aggregates that predict a gradual decrease over 10s of km for typical (5-15 °C/km) mantle geotherms. Secondly, lab-based electrical conductivity experiments struggle to reflect the increase in conductivity observed in the asthenosphere (Yoshino et al., 2017).

The magnitude of the viscosity drop is inconsistent with single-crystal plasticity and creep models. Experimental flow laws based on single-crystal plasticity predict a much stronger lithosphere than is observed geophysically across a large range of depths (Raterron et al., 2019; Tielke et al., 2016). Results at both low and high pressures show that grain boundaries weaken olivine aggregates compared to single-crystal estimations, in line with viscosities observed through the mantle up to 400 km. However, this yields prediction of a gradual decrease in strength with increasing temperature and depth, which cannot account for the sharp drop in viscosity from the lithosphere to the asthenosphere. This implies that grain boundary processes are essential for deformation across much of the upper mantle and that the discontinuity at the LAB is associated with changes in the grain boundary structure and network.

Miyazaki et al. (2013) observe a step change in the microstructure of forsterite + diopside (Fo + Di) aggregates, where below 1250 °C, grains are equiaxed, and above 1250 °C, grains are tabular/elongated.

This change in microstructure is coupled with the emergence of a crystallographically preferred orientation (CPO) during deformation experiments performed across this temperature threshold. Maruyama & Hiraga (2017a, 2017b) show that these samples with tabular grains exhibit lower viscosity during diffusion-assisted grain-boundary sliding (GBS) creep, which operates at lower-lithospheric conditions. These works identify enhanced deformation rates at $T>0.84-0.92T_s$ (where $T_s$ is solidus temperature). For typical mantle rocks this ranges from 1150 to 1540 °C and coincides with the depth range where weakening into the asthenosphere is observed. It is speculated that 'grain boundary disordering' occurs across this temperature threshold but direct evidence is lacking.

Secondly, experimental electrical conductivity measurements on olivine aggregates are inconsistent with with large-scale geophysical observations of the conductivity at the LAB using magnetotellurics. Beneath the Slave craton, a transition in conductivity from 0.0033 S/m to 0.0133 S/m is observed at ~210 km (~1300 °C) (Bettac et al., 2023b; Jones et al., 2003) and an order-of-magnitude drop in resistance is seen between 150-200 km beneath the Kaapvaal craton (Evans et al., 2011). In laboratory experiments comparing conductivity in olivine single crystals vs. aggregates between 3.5-14.9 GPa and 1030-1830 °C, the activation enthalpy for conduction increases across a temperature threshold in single crystals but remains constant in aggregates (Yoshino et al., 2017). This increase causes a shoulder in conductivity plots. The temperature of this shoulder increases with increasing pressure from 1265 °C at 3.4 GPa to 1540 °C at 9.6 GPa. The absence of a shoulder in experimental conductivity measurements on polycrystalline olivine is inconsistent with observations of conductivity variations in the dry mantle, as measured by magnetotelluric methods.

In this work we present, for the first time, experimental evidence of a complexion transition in forsterite that explains the two open questions introduced. A temperature-dependent complexion transition occurs between 1290 °C and 1390 °C, altering the grain boundary energy anisotropy and changing the dominant grain boundary planes in a forsterite polycrystal. This reversible transition provides a physical explanation for abrupt changes in viscosity and conductivity observed beneath cratons and unifies previous experimental observations of grain-shape and boundary-property changes across the $T>0.9\ T_s$ threshold.

**Results**

Forsterite + 2 % enstatite aggregates sintered at temperatures between 1150 °C and 1390 °C were investigated by EBSD to construct the forsterite grain boundary plane distribution (GBPD). The evolution of the GBPD with increasing sintering temperature is shown in Figure 1. The GBPD is the *distribution of grain boundary planes*, irrespective of disorientation, in other words the average habit crystal. The plots are colour-coded in multiples of random distribution (MRD) and show how likely a given plane is to be observed compared to a random distribution. Samples sintered between 1150 °C and 1290 °C have the same GBPD, where the most commonly occurring grain boundary planes are close to perpendicular to the *c-axis,* with maxima occurring between {014} and {001} (termed *c-type*). The least common grain boundary planes are clustered close to {100}. Notably, a similar GBPD was observed by Marquardt & Faul (2018) for a considerably larger grained sample (~20 µm grain size) for similar synthesis conditions. At 1390 °C the GBPD is *markedly* different from the GBPD at lower temperatures. The anisotropy of the distribution (difference between maximum and minimum MRD) increases from 1.13 at 1290 °C to 1.81 at 1390 °C. While the least common grain boundary planes remain {100}, the most common grain boundary planes transition to {010} (termed *b-type*). The occurrence of the most frequently observed grain boundary planes increases from ~1.5 to 2.13 times the number expected in a random distribution.

To test if this change in grain boundary population is reversible, a piece of the low-temperature sample (LT_1150) was annealed above the temperature threshold at 1390 °C for 4 h, and a piece of the high-temperature sample (HT_1390) below the temperature threshold at 1290 °C for 34 h (experimental conditions summarised in Table 1 S.I.). After annealing, the average grain size of both samples increased: LT_1150 from 0.6 µm to 4.53 µm and HT_1390 from 6.64 to 6.90 µm. After annealing, the GBPD of LT_1150 transformed from *c-type* to *b-type*. The most frequently observed grain boundary planes changed from (021) to {010} and the frequency of the most common planes increased from 1.41 to 1.74 MRD. The difference plot of the GBPD before and after annealing in Figure 2a shows the increase of {010} grain boundary planes and decrease in {001} and {100}. The changes to the GBPD of sample HT_1390 after annealing at 1290 °C were more subtle. The most common grain boundary planes remained {010}, but the frequency with which they are observed decreased from 2.13 to 1.84 MRD. The difference plot in Figure 2b shows the decrease in occurrence of these planes and the increase in frequency of grain boundary planes close to {001}.

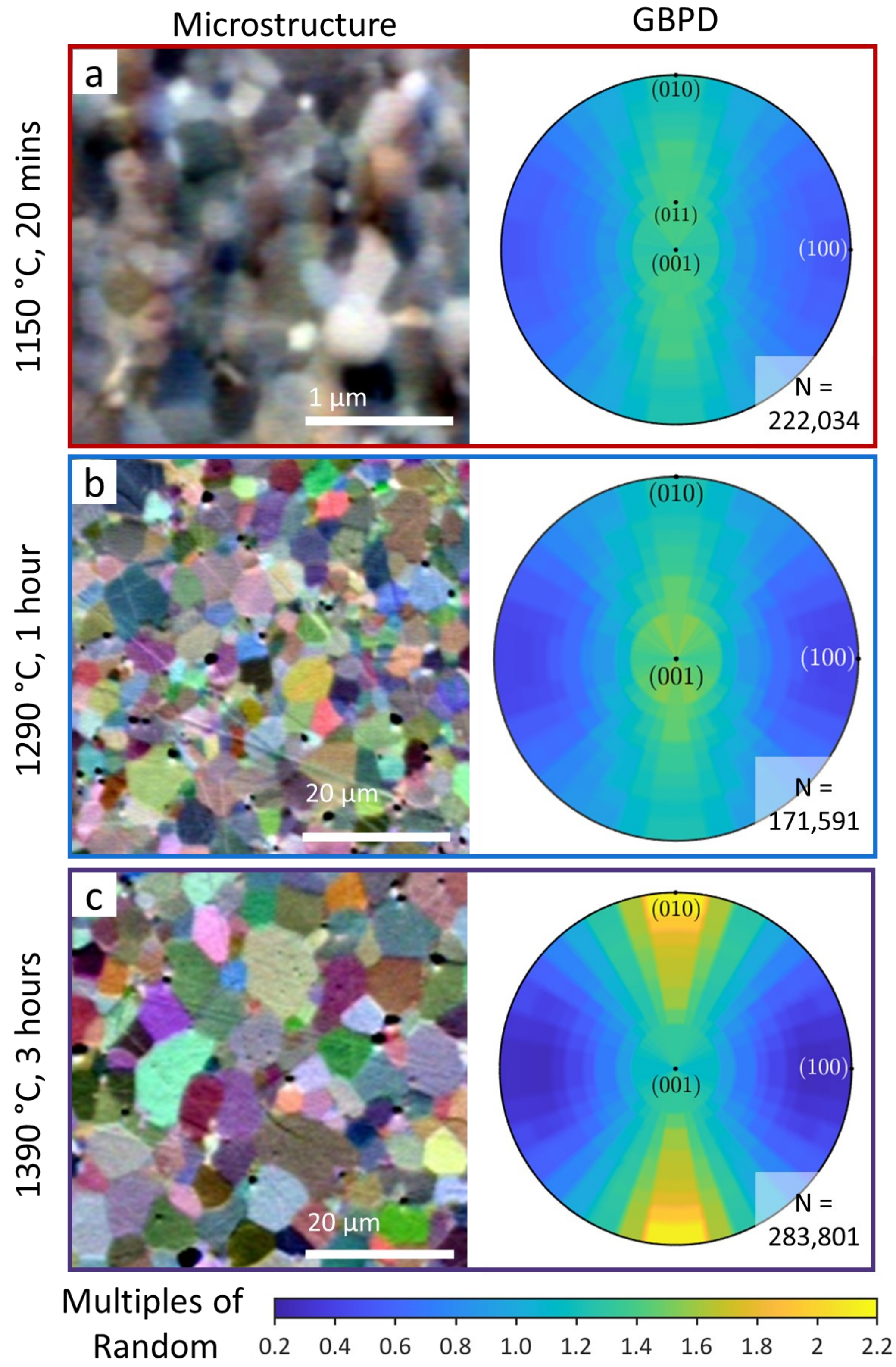

**Figure 1:** Microstructure and GBPD of samples a) LT_1150, b) LT_1290 and c) HT_1390. Column 1: Forescattered electron (FSE) image of sample microstructure. Column 2: GBPDs obtained from 2D EBSD maps using stereological methods. GBPDs for all samples are plotted on the same colour scale from 0.2-2.2 MRD and have number of segments used for construction (N) indicated.

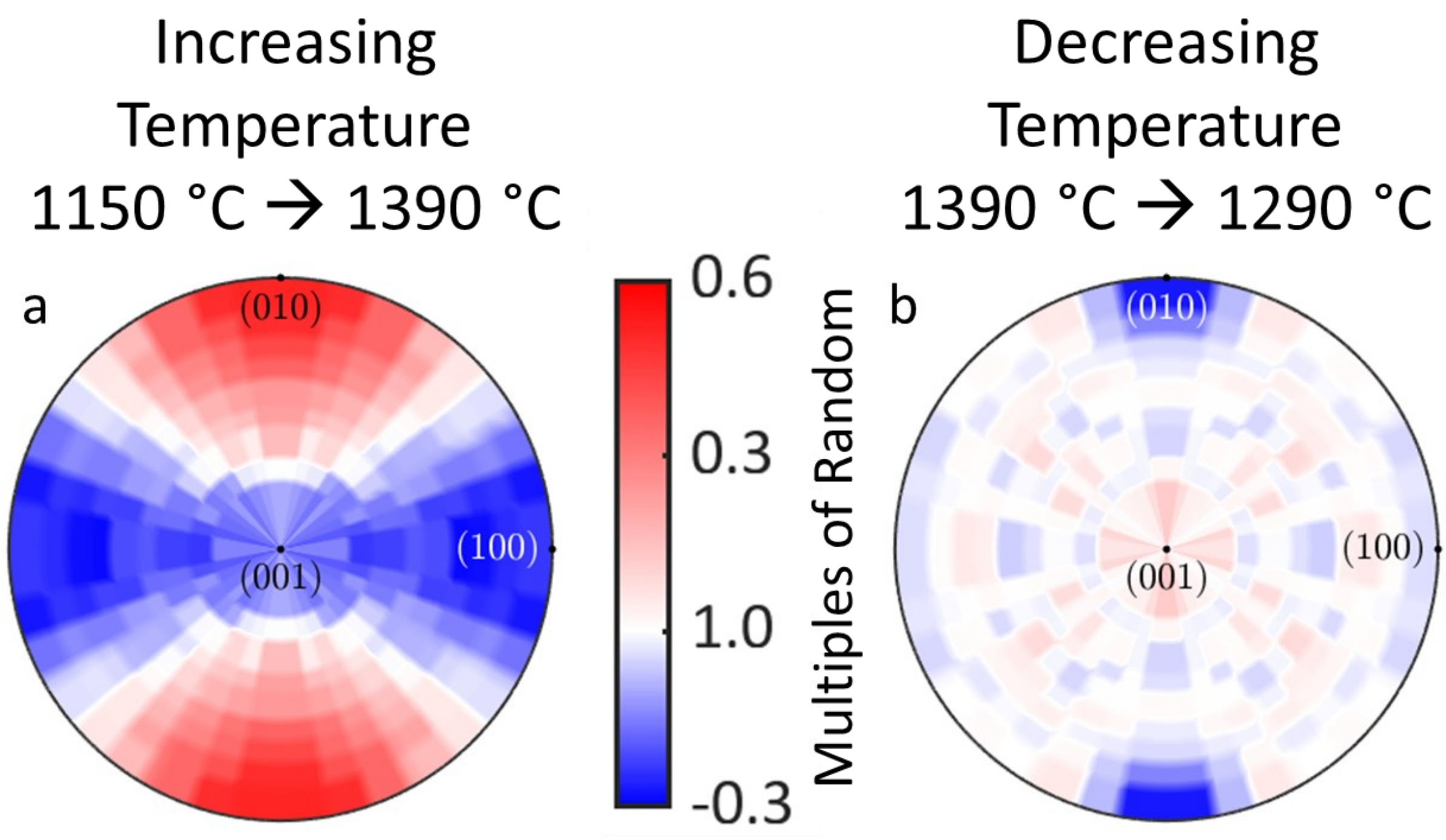

**Figure 2:** Stereographic projections showing changes in the GBPD before and after annealing experiments. a) LT_1150 synthesised at 1150 °C, annealed at 1390 °C for 4h, b) K HT_1390 at 1390 °C, annealed at 1290 °C for 34 h. Red indicates an increase in MRD value after annealing, blue a decrease. An extended figure showing GBPD of annealed samples is available in the extended data.

**Discussion**

The GBPD can be thought of as an approximation of the average habit crystal shape in an aggregate, with an inverse relationship to the grain boundary energy distribution (GBED) (Holm et al., 2011; Rohrer, 2011; Saylor, El Dasher, et al., 2004; Saylor et al., 2003b, 2003a). The most common grain boundary planes in the GBPD and the faces with the largest area in a habit crystal have the lowest energy. Experimentally and in nature, olivine crystal habits have been directly observed to be dominated by b-planes. However, the average crystal shape in an olivine aggregate is not known and it is commonly assumed that there is no strong grain boundary energy anisotropy (Cooper & Kohlstedt, 1982; Duyster & Stöckhert, 2001), qualitatively argued by an equiaxed grain structure with low aspect ratios (Vernon, 2018). In contrast, the GBPD in this study indicate a clear variation in grain-boundary energy and that the lowest-energy grain-boundary planes change across a temperature range of 1290 °C-1390 °C. Below this threshold, the most frequent boundaries are {0kl}-type, *c-type* GBPD, and above it, {010} boundaries dominate- a *b-type* GBPD. This change reflects a reordering of the most stable interfacial structures or a grain boundary complexion transition.

Annealing experiments demonstrate that the grain boundary complexion transition is reversible and independent of grain size or chemistry. In the as-synthesised samples, the GBPD is *c-type* at 1290 °C and below and *b-type* at 1390 °C (Figure 1, Extended data for all samples analysed). The sample synthesised at 1150 °C and annealed at 1390 °C exhibited the high-temperature GBPD, whereas the sample sintered at 1390 °C and annealed at 1290 °C reverted to the low-temperature GBPD. The difference plots in Figure 2 highlight the changes to the GBPDs after annealing. Due to the larger starting grain size of HT_1390 and the annealing temperature being below the initial sintering temperature, there is less driving force for grain growth over the duration of these annealing experiments. This results in slower grain growth and consequently less grain boundary migration and a less-pronounced change in the GBPD. Nevertheless, grain growth is sufficient to allow grain boundaries to adjust their orientation, thereby allowing the GBPD to reflect the new low-energy complexion. It should be noted that at geological timescales, we expect microstructures to equilibrate. All samples have the same chemistry and phase assemblage, and the high-temperature complexion could be transformed to the low-temperature one during normal grain growth. This demonstrates that the origin of the different grain boundary distributions is a temperature-dependent transition in grain boundary complexion. While GBPD changes have been employed to describe complexion transitions in doped $Al_2O_3$ and $Y_2O_3$ samples heated in different complexion-stability fields (Bojarski et al., 2012; Dillon et al., 2010), here we report the first experimental observation of a complexion transition in a rock-forming silicate.

The change in the most common grain boundary plane across a temperature threshold indicates a change in the relative grain boundary energies and, consequently, a change in the grain boundary structures between 1290 and 1390 °C (see Rohrer, 2016). The low-temperature GBPDs observed in this study have a weak anisotropy (0.9-1.2 MRD), while the high-temperature GBPD displays a more-substantial anisotropy (1.86 MRD). This indicates that the range of grain boundary energies is smaller at 1290 °C and below than at 1390 °C. The prominence of *b-planes* at high temperatures indicates that their energy has a weaker temperature dependence than other grain boundary planes in this system.

Geophysical observations reveal a pronounced rheological discontinuity at the base of cratonic lithospheres. Classical flow laws for olivine cannot reproduce the sharp rheological transition in the absence of melt or hydration. The grain boundary complexion transition documented here provides a physical explanation for this discrepancy: a temperature-driven restructuring of the grain boundary network that dramatically reduces polycrystal viscosity.

As the grain boundary structure changes and grain boundary energies become more anisotropic, the prominence of certain planes increases, and their properties come to dominate creep behaviour. Dislocation glide in olivine is generally confined to b-planes and there is evidence that GBS favours the formation of b-planes at grain boundaries (Ferreira et al., 2021). We observe a complexion transition above ~1290 °C that stabilises b-plane boundaries, thereby increasing the area available for slip. This facilitates grain-boundary sliding and sharply reduces bulk strength. This step change in the grain boundary network occurs in the temperature range of the viscosity drop beneath cratons and provides the missing piece as to why a gradual decrease in viscosity is predicted by classical flow laws but not observed in nature (Raterron et al., 2019). Independent deformation experiments support this interpretation. Miyazaki et al. (2013) observed a transition from a very weak E-type to A-type CPO at 1250 °C in diffusion creep experiments of Fo + Di aggregates synthesised in the same way as the aggregates in this study. A change in grain morphology was also noted, with tabular/elongated olivine grains, bounded by 'low-index' planes, being observed in samples sintered at T>1250 °C. These low-index planes have been shown to have ~3 times lower viscosity than general boundaries (Maruyama & Hiraga, 2017b, 2017a) during GBS. These results demonstrate that this change in creep behaviour arises from a grain boundary complexion transition in forsterite, providing the missing microstructural link between laboratory rheology and geophysical viscosity profiles.

This complexion transition can also explain discrepancies between magnetotelluric and lab-based conductivity measurements. Beneath cratons, magnetotelluric data show an order-of-magnitude increase in conductivity at depths of 180-230 km, yet equivalent jumps are not consistently observed in laboratory experiments. Often, no change or a small shoulder is seen in the conductivity of polycrystalline aggregates (Omura et al., 1989; ten Grotenhuis et al., 2004; Tyburczy & Roberts, 1990; Yoshino et al., 2017). Experiments that observe a shoulder in the grain boundary conductivity generally involve longer equilibration times, or are more chemically complex, thereby lowering the solidus temperature and hence complexion transition temperature.

Complexion transitions occur at the atomic scale, where restructuring at grain boundaries happens rapidly and without visible microstructural change. Related, microstructural adjustments governed by surface energy minimisation after an atomistic complexion transition require significant grain growth before boundary planes can change orientation and relative area. This requires sufficient driving force and time, so whether a conductivity shoulder is observed depends on the grain-boundary network's ability to equilibrate at a given homologous temperature. In chemically pure, short-term laboratory experiments, the microstructural changes are kinetically inhibited. In contrast, in Fe-bearing or older, natural systems the microstructure resulting from a complexion transition is fully expressed, resulting in the conductivity discontinuities seen beneath cratons.

Independent evidence links the increase in conductivity to the presence of *b-type* planes. Pommier et al. (2018) reported that samples deformed to the highest strain in shear experiments exhibited the highest conductivity. These samples had an A-type fabric with *a-axes* aligned parallel to the shear direction and *b-axes* perpendicular. Ferreira et al. (2021) calculated the GBPD in sheared samples that display the same fabric and found that at high strains the most common grain boundary planes are *b-type*, indicating that the highly conductive grain boundary planes in Pommier's experiments are *b-type*. Together, these studies show that different grain boundary planes have different conductivities. Over geological timescales, where microstructural equilibration is expected to be complete, the conductivity beneath cratons reflects the increased area fraction of *b-type* boundaries after the complexion transition.

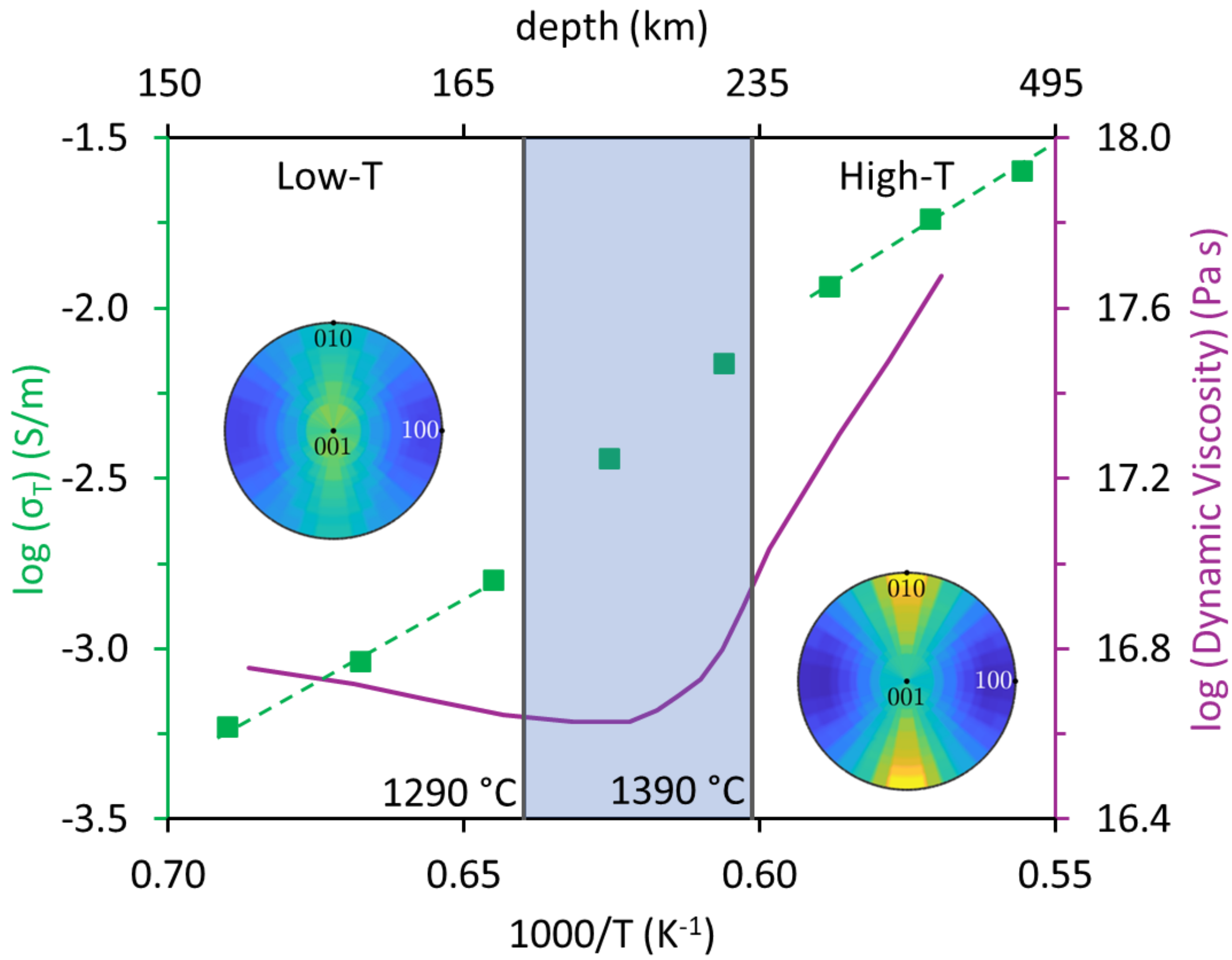

**Figure 3:** Diagram illustrating complexion stability fields. Variation in olivine aggregate dynamic viscosity for a continental geotherm from Raterron et al. (2019) and total forsterite aggregate conductivity ($\sigma_T$) from Yoshino et al. (2017) with temperature.

We have demonstrated, for the first time, a temperature dependent complexion transition in forsterite. Between 1290 and 1390 °C, the grain boundary network transitions from c-plane dominated to b-plane dominated. This change in the grain boundary network explains the sharp viscosity drop across the lithosphere-asthenosphere boundary where grain boundary processes govern creep and deformation. It can also explain the conductivity change across this boundary beneath cratons, summarised in Figure 3. Given the grain-boundary network's integral role in melt percolation, creep behaviour, and electrical conductivity, changes to it could have far-reaching implications. Our findings suggest that grain boundary complexion transitions can modulate mantle viscosity, seismic anisotropy, and conductivity - linking interface-scale thermodynamics to macroscopic Earth structure.

**Methods**

**Specimens:** Five synthetic 0.98 $Mg_2SiO_4$ + 0.02 $MgSiO_3$ aggregates with grain sizes from 600 nm to 6.6 μm were synthesised at the Earthquake Research Institute at The University of Tokyo by the method described in Koizumi et al. (2010). LT_1150 was densified using Spark Plasma Sintering (SPS) while all other samples were sintered in a vacuum furnace. Table 1 S.I summarises the synthesis conditions and subsequent heat treatments of the datasets reported.

Annealing experiments were performed to assess the stability of the GBPD after sample synthesis. A piece of sample LT_1150 (sintered at 1150 °C) was annealed in a vacuum furnace at 1390 °C for 3 h in an alumina crucible on a bed of $Mg_2SiO_4$ powder (henceforth referred to as LT_1150a); a piece of HT_1390 was annealed at 1290 °C for 34 h in the same crucible set-up (henceforth referred to as sample HT_1390a).

Pellets were sectioned through-thickness and mechanically polished to a mirror finish with a final step of vibratory polishing with colloidal silica. To prevent charging during analysis, samples were coated with a 2 nm carbon film.

**Electron Microscopy:** FSE images were acquired on a TFS FEI Quanta 650 SEM at 20 kV with a Bruker eFlash HD detector with ARGUS imaging system. EBSD data of samples K3-6 were collected at 20 kV accelerating voltage with a 6.4 nA current using a TFS Scios Dual Beam FIB equipped with an Edax Velocity detector. A Zeiss Sigma FEG SEM with an Edax Clarity Detector operating at 10 kV with a 60 μm aperture was used to analyse sample LT_1150. Maps were collected encompassing over 50,000 grains per sample using the conditions summarised in Table 2 S.I.. The step size was chosen to be at least 1/10$^{th}$ of the olivine grain size. The near surface and bulk of K4 were mapped separately to assess changes to the GBPD with grain size within the same sample. The EBSPs were indexed during data collection using the orthorhombic space group 62, Pbnm, lattice parameters a=4.75 Å, b=10.2 Å, c=5.98 Å for forsterite and the orthorhombic space group 62 Pnma, lattice parameters a=18.24 Å, b=8.83 Å, c=5.18 Å for enstatite. Note here that crystallographic axis definition for forsterite can also be Pnma. Care should be taken when comparing results from other sources to check consistency of the unit cell definition.

To extract the grain boundary segments used for GBPD construction, the EBSD data were post-processed using OIM TSL Analysis v8. Example maps from each step are shown in Figure 1 S.I. Firstly, points with a confidence index <0.1 were reindexed using the Hough transform method and NPAR pattern averaging (Wright et al., 2015). Secondly, mis-indexing related to pseudosymmetry (PS) caused by the quasi-hexagonal oxygen lattice parallel to the *a*-axis ([100]) of $Mg_2SiO_4$ (Poirier, 1975) was addressed (Figure 2 S.I.). Finally, grain dilation was performed to assign mis-indexed points at grain boundaries to one of the neighbouring grains and enable grain boundary traces to be reconstructed. Further details on data processing can be found in the methods of Marquardt et al. (2015) and Ferreira et al. (2021). All processing and reindexing parameters are summarised in Table 2 S.I. The grain boundary segments were reconstructed in OIM with a maximum deviation of 2 steps from the edge of the grain identified. These segments were exported and sorted so only boundaries between $Mg_2SiO_4$ grains with a size greater than a specified size were used for constructing the grain size distributions, disorientation distributions and GBPD. The Mtex toolbox in Matlab was used to calculate the orientation distribution texture indices and function (ODF) with a single point per grain (Mainprice et al., 2015). This is necessary as the absence of strong texture is a prerequisite for the application of stereological methods to a 2D EBSD map for the calculation of GBPD and GBCD(Randle & Davies, 2002). The exported segments were processed using the programmes developed at CMU by Rohrer et al. (2004) to calculate the disorientation and grain boundary plane distributions.

The grain size was taken as the equal area diameter of a circle with the same area as each grain in an EBSD map. These were used to construct the grain size distributions and determine the average grain size of the samples.

**Extended Data**

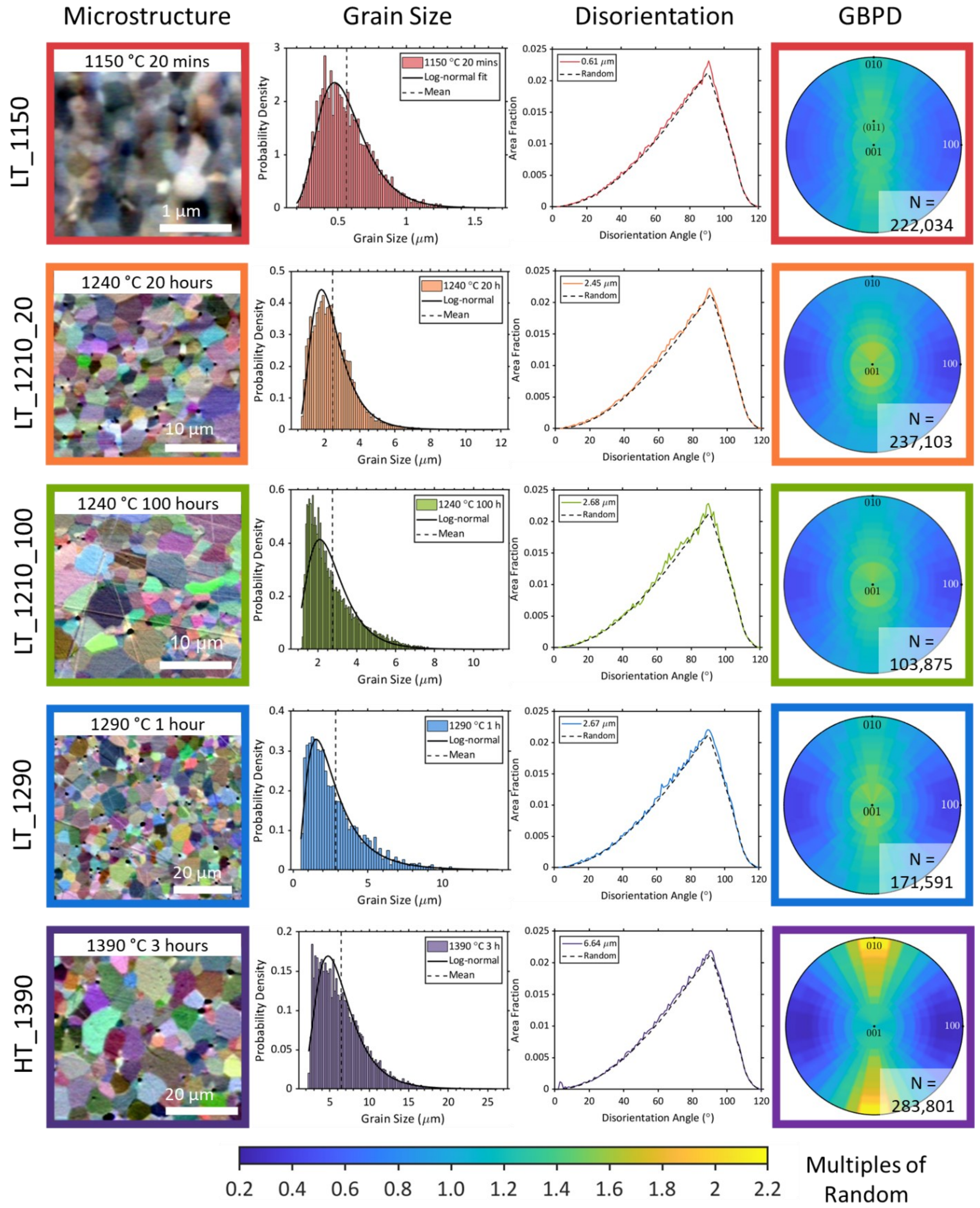

**Figure 1 Extended Data:** Microstructure, grain size distribution, disorientation distribution and GBPD for samples sintered between 1150-1390 °C.

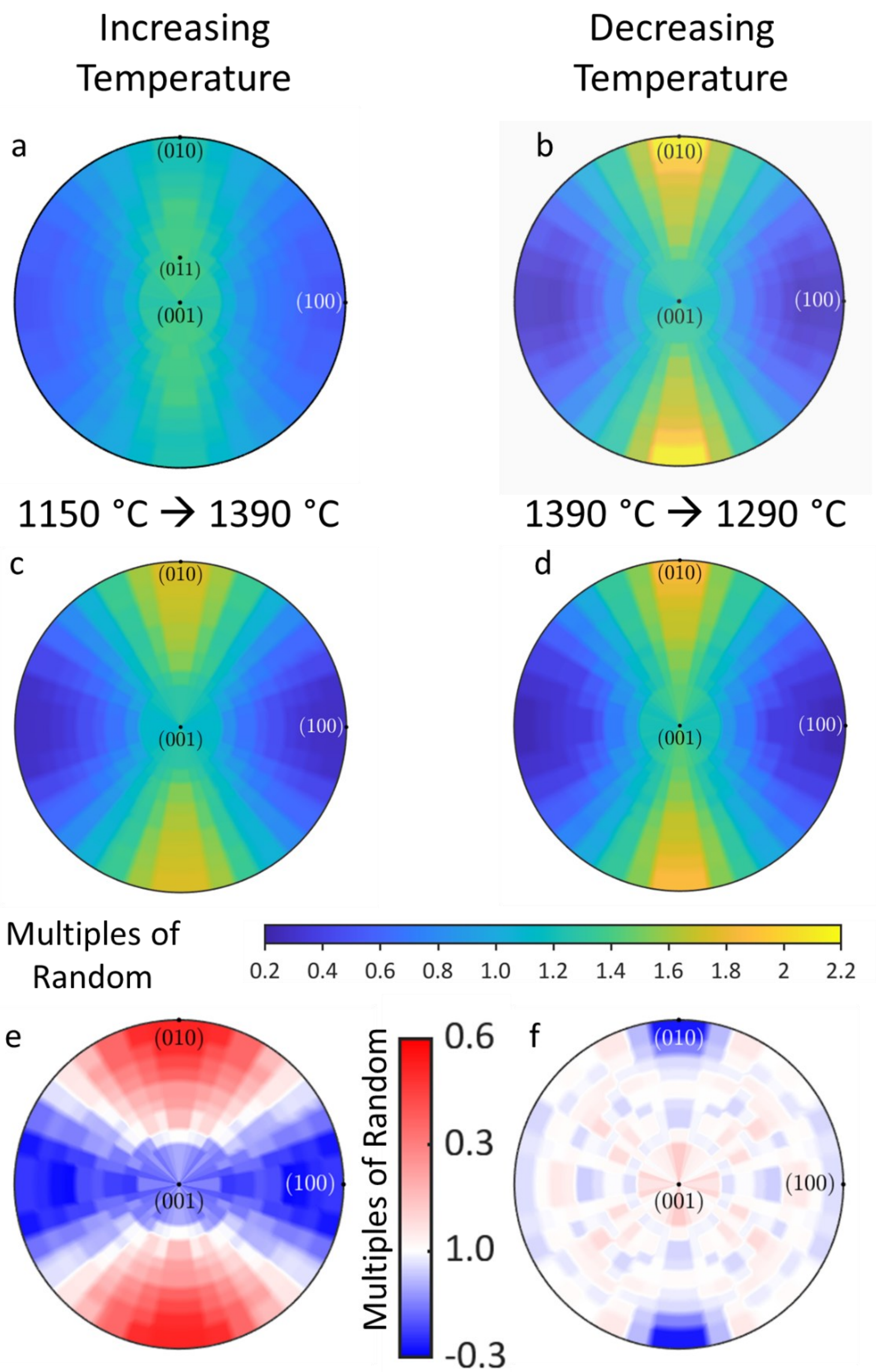

**Figure 2 Extended Data:** Complexion reversibility annealing experiments. GBPD of as-synthesised samples, a) LT_1150 and b) HT_1390. GBPD after annealing c) LT_1150 at 1390 °C for 4h and d) HT_1390 at 1290 °C for 34h. All GBPD are plotted on a scale from 0.2-2.2 MRD. Stereographic projections showing changes in the GBPD before and after annealing experiments. e) increasing temperature experiment, f) decreasing temperature experiment. Red indicates an increase in MRD value after annealing, blue a decrease.